\\

Title: Melioration of the radiocesium contaminated land

Authors: I. E. Epifanova, E. G. Tertyshnik

A method is described of radiocesium fixation in soils contaminated by this radionuclide. To immobilize radiocesium, the soil surface is treated with aqueous hexacyanoferrate solution of alkaline metals. It has been experimentally shown that application of $K_4[Fe(CN)_6] \cdot 3H_2O$ at a rate of 1,3g/kg soil reduces the fraction of exchangeable $^{137}Cs$ 100-fold (100 times). The method is effective for the plots where contamination is concentrated in the top 1 – 2 cm soil layer.

Comment: 4 pages, including 1 table
Subjects: Nuclear Experiment (nuc-ex), Geophysics (physics.geo-ph)
\\

**Melioration of the radiocesium contaminated land**

I. E. Epifanova (Russian Institute of Agricultural Radiology and Agroecology),

E. G. Tertyshnik (Research and Production Association "Typhoon", Obninsk, Kaluga region, Russian Federation).

In the aftermath of nuclear accidents agricultural lands are affected by $^{137}$Cs and $^{134}$Cs contamination.

To reduce radiocesium transfer from the contaminated soils to yield, soils are treated by chemicals, as well as solutions of ferric chloride, alkaline metal phosphates to immobilize radionuclides in the soil and prevent their uptake by plants [1 – 3]. The limitations of these approaches are changes in the soil acidity and low effectiveness of radiocesium fixation.

An approach is described [3] of soil treatment by potassium compounds (potassium fertilizers) and combined fertilizers which contain potassium to reduce the soil content of mobile (exchangeable) radiocesium. According to this approach, soil samples were artificially contaminated with $^{137}$Cs by way of their treating by solutions that contained the given radionuclide, aged for 1, 2 and 5 years and the fraction of exchangeable $^{137}$Cs (nonfixed by soil and, consequently, available for plants) was determined depending on the conditions of mineral fertilizers application. The methodology for exchangeable $^{137}$Cs determination is described in [4]. The results obtained demonstrate [3] that the application of potassium fertilizers alone and combined use of potassium, phosphate and nitrogen fertilizers reduces the content of exchangeable $^{137}$Cs at the most 4 times, and some combinations of fertilizers added to soils rich in potassium, on the contrary, result in increase in the fraction of exchangeable radiocesium in the soil compared to the control and corresponding increase in $^{137}$Cs uptake by vegetable crops.

To improve the efficiency of radiocesium fixation in soils it has been suggested [5] that soil contaminated by radiocesium was treated with aqueous hexacyanoferrate (HCF) solution of alkaline metals, for instance, aqueous solution of potassium hexacyanoferrate (KHCF). Since any type of soil contains metals such as calcium, magnesium, iron, aluminium, titanium, zinc, copper, cobalt, etc., the applied to soil complex anions $[Fe(CN)_6]^{4-}$ or $[Fe(CN)_6]^{3-}$ form with these metals double or single HCFs, for example, $K_2Ca[Fe(CN)_6]$, $Ca_2[Fe(CN)_6]$, $KFe[Fe(CN)_6]$, etc. Most of these HCFs are poorly soluble compounds. All formed HCFs selectively and effectively bind cesium and radiocesium ions. In this case poorly soluble double salts of $Cs_2Ca[Fe(CN)_6]$ and $Cs Fe[Fe(CN)_6]$ type are formed. As a result, in the treated soil the number of atoms (ions) of cesium and radiocesium capable of ion exchange and plant uptake is reduced. The resulting HCFs such as $Cs_nMe_m[Fe(CN)_6]_k$ are radiation resistant, chemically resistant to acids and concentrated solutions of neutral salts; alkaline solutions destroy hexacyanoferrates (except for cobalt HCF). High pH values are, however, not typical for soil solutions. Due to HCF property to selectively fix cesium and considering the fact that concentration of stable cesium nuclides in soil is extremely negligible (on average $5 \cdot 10^{-4}$ %), relatively small amounts of HCF solution are needed to be applied to reduce the content of exchangeable radiocesium in the soil.

To demonstrate the effectiveness of radiocesium fixation in soil using the proposed method, a plot of virgin soil (dry meadow) within the 30 km zone near the Chernobyl NPP was investigated in 1988. By means of soil sampling with subsequent soil division into layers and measurement of gamma-emitting radionuclides in each layer it was found that the depth of radionuclide contaminated soil layer on this plot was 1-2 cm. On the plot, samples of 1 cm top soil layer were collected, 100 g mass each. In each sample, $^{137}Cs$ content was measured using a gamma-spectrometric device equipped with the HPGe-detector. The surface of the soil samples (surface area – 100 cm$^2$) was treated with 30 ml potassium HFC solution at a concentration of 0,001 M and dried at room temperature till the air-dried condition. The fraction of mobile (nonfixed in the soil) $^{137}Cs$ in the treated and control (untreated with KHCF) soil sample was determined using the well-known methodology [4]: poured 200 ml 1-normal ammonium acetate solution into a soil sample, mixed thoroughly for 10 minutes, infiltrated and treated again with a new portion of ammonium acetate. Both portions of the filtered material were amalgamated and $^{137}Cs$ content was measured with a gamma-spectrometer. The fraction of radionuclides extractable with 1-normal ammonium acetate solution is known to define the relative amount of mobile radionuclides which can escape from soil to plants.

It has been found that in the control soil sample (untreated with KHCF) the share of mobile (exchangeable) $^{137}Cs$ amounted to 4%. In the soil sample treated with 0,001 M solution of $K_4[Fe(CN)_6] \cdot 3H_2O$ , the share of exchangeable $^{137}Cs$ was 0,12%, i.e., dropped 33 times compared to the control.

The table below summarizes results of the experiment when soil samples were treated with the equal portions of solution (30 ml) containing different amounts of KHCF. The residence time of $^{137}Cs$ in soil is 2 years (since the moment of the Chernobyl accident). The average specific activity of the used soil samples caused by $^{137}Cs$ is $5 \cdot 10^5$ Bq/kg. As seen from the table, the optimal rate of $K_4[Fe(CN)_6] \cdot 3H_2O$ used for $^{137}Cs$ fixation in the soil varies from 1,3 to 1,9 g / kg soil. At a KHCF rate of 0,063 g/kg treated soil, the fraction of mobile (exchangeable) radiocesium is rather high – up to 2,8%. Consequently, the effectiveness of KHCF application at this concentration is low. At a KHCF rate above 1,9 g/kg soil (above 247 kg/ha), the share of exchangeable $^{137}$Cs in soil remains at a level of 0,04 %. Consequently, the use of KHCF above this value is not advisable.

T a b l e. Effects of soil treatment with $K_4[Fe(CN)_6] \cdot 3H_2O$ solution (molecular weight 422) per fraction exchangeable $^{137}$Cs in soil

| No. | $K_4[Fe(CN)_6] \cdot 3H_2O$ molar concentration (at 300 ml solution per kg soil) | Substance amount, g/kg soil | Rate of applied substance, kg/ha | Share of exchangeable $^{137}$Cs, % |
|---|---|---|---|---|
| 1 | Untreated sample (control) | - | - | 4,0 |
| 2 | 0,0005 M | 0,063 | 8,19* | 2,8 |
| 3 | 0,001 M | 0,127 | 16,5* | 0,12 |
| 4 | 0,005 M | 0,634 | 82,4* | 0,06 |
| 5 | 0,01 M | 1,267 | 165* | 0,038 |
| 6 | 0,015 M | 1,900 | 247* | 0,036 |
| 7 | 0,03 M | 3,800 | 494* | 0,037 |

\* — estimated for the 1 cm contaminated soil layer, soil density is taken to be 1,3 kg/dm$^3$. Contaminated soil mass per 1 ha is 130 t.

According to the table data, for treat a land plot of an area of 1 ha 39 tons of solution are needed to be distributed uniformly by means of spray or splash application. It is evident that this amount of solution can be reduced 5— 10 times (to a wise volume) while the mass of applied KHCF keep the former size (165 kg/ha), since it is possible due to high KHCF solubility in water.

Radioactive contamination of soil following nuclear weapons tests or accidents at nuclear fuel cycle facilities occurs through the deposition of radionuclides on the soil surface. The layer of contaminated soil at the moment of contamination is very thin (film contamination). Within several years, due to radionuclide migration deeper into the soil, the contaminated layer depth can extend to several centimeters. Therefore, when estimating the HCF amount to be applied according to the proposed method, the depth of radiocesium contaminated soil layer needs to be taken into account. The use of the given method for radiocesium fixation in areas where contamination is distributed throughout the entire arable layer seems to be inappropriate.

Since HCF solution is applied to the soil surface, with depth concentration of complex anions in solution is decreasing and conditions of radiocesium fixation deteriorate. To improve the situation, it is advisable to add into the solution ballast salts of Na or K, for example, sodium chloride. In this case chlorine anions are fixed in the soil layers, and since anion exchange capacity of soil is low, absorption of complex anions $[Fe(CN)_6]^{4-}$ or $[Fe(CN)_6]^{3-}$ in soil layers declines.

**Conclusion**

A distinctive feature of the proposed approach to radiocesium location in soils is application to the soil surface of aqueous HCF solutions of alkaline metals which, when interacting with soil microelements, form insoluble compounds that selectively absorb cesium ions.

Simulated experiments have demonstrated that treatment of $^{137}Cs$ contaminated soil by aqueous solution of potassium hexacyanoferrate (at a rate of 1,3 g/kg soil) results in a 100-fold decrease in the fraction of exchangeable $^{137}Cs$.